# Metallic and insulating interfaces of amorphous $SrTiO_3$-based oxide heterostructures

Yunzhong Chen,*[1] Nini Pryds,[1] Josée E. Kleibeuker,[2] Gertjan Koster,[2] Jirong Sun,[3] Eugen Stamate,[4] Baogen Shen,[3] Guus Rijnders,[2] and Søren Linderoth[1]

[1] Fuel Cells and Solid State Chemistry Division, Risø National Laboratory for Sustainable Energy, Technical University of Denmark, DK-4000 Roskilde, Denmark

[2] Faculty of Science and Technology and MESA+ Institute for Nanotechnology, University of Twente, 7500 AE Enschede, The Netherlands.

[3] Institute of Physics, Chinese Academy of Sciences and Beijing National Laboratory for Condensed Matter Physics, 100190 Beijing, China.

[4] Plasma Physics and Technology Programme, Risø National Laboratory for Sustainable Energy, Technical University of Denmark, DK-4000 Roskilde, Denmark.

Emails: yunc@risoe.dtu.dk, nipr@risoe.dtu.dk, J.E.Kleibeuker@utwente.nl, g.koster@utwente.nl, jrsun@g203.iphy.ac.cn, eust@risoe.dtu.dk, shenbg@g203.iphy.ac.cn, a.j.h.m.rijnders@utwente.nl, sqli@risoe.dtu.dk.

Nano Letters

---

*     Corresponding Author. Email:yunc@risoe.dtu.dk; Phone: +45 4677 5614; Fax: +45 4677 5858.

**ABSTRACT**: The conductance confined at the interface of complex oxide heterostructures provides new opportunities to explore nanoelectronic as well as nanoionic devices. Herein we show that metallic interfaces can be realized in SrTiO$_3$-based heterostructures with various insulating overlayers of amorphous LaAlO$_3$, SrTiO$_3$ and yttria-stabilized zirconia films. On the other hand, samples of amorphous La$_{7/8}$Sr$_{1/8}$MnO$_3$ films on SrTiO$_3$ substrates remain insulating. The interfacial conductivity results from the formation of oxygen vacancies near the interface, suggesting that the redox reactions on the surface of SrTiO$_3$ substrates play an important role.

Strontium titanate is a prototype wide-gap insulator with a perovskite structure. Due to the structural compatibility, SrTiO$_3$ (STO) has been widely used as a substrate material for the growth of, among others, high-temperature-superconducting cuprates, colossal magnetoresistive manganites, and multiferroics. Particularly, the accessibility of regular singly TiO$_2$-terminated STO substrates[1,2] triggered the possibility to fabricate complex oxide heterostructures with atomic-scale-flat interfaces through controlled film growth methods, such as pulsed laser deposition (PLD) equipped with reflection high energy electron diffraction (RHEED).[3] Recently, a broad spectrum of interesting properties, such as a quasi-two dimensional electron gas (q-2DEG),[4,5] magnetism,[6] charge writing,[7] resistance switching,[8] giant thermoelectric effect,[9] and colossal ionic conductivity[10] have been observed in various oxide heterostructures based on STO substrates. These conductance-related interfacial functionalities offer potential applications in oxide electronics,[11] thermoelectric materials, and solid oxide fuel cells. To date, the origins and the intrinsic mechanisms for these emerging properties have been studied intensively.[12] Besides electronic reconstruction near the interface, which is found to play an important role in determining the interfacial properties,[4-7,11,13] mechanisms such as ion transfer across the interface and formation of defects have also been identified as important factors for the transport properties.[14, 15] However, the chemical driving forces leading to compositional changes across the interface are rarely explored.

Here, metallic conducting interfaces are observed in STO-based heterostructures with various overlayers of amorphous LaAlO$_3$ (LAO), STO, and yttria-stabilized zirconia (YSZ) films. Whereas, an insulating heterointerface is found when the overlayer is an amorphous La$_{7/8}$Sr$_{1/8}$MnO$_3$ (LSMO) film. We present evidence that the interfacial conductivity results from the oxygen vacancies on the STO substrate side due to the exposure of the SrTiO$_3$ substrate surface to reactive species of film growth. Although the energy of the arriving species has been suggested to be responsible for the creation of defects,[15-17] the chemical reactivity of these species at the STO substrate surface has not been considered yet, to our knowledge. With our results we show that the latter mechanism is an important source for the creation of mobile charge carriers in STO-based oxide heterostructures.

We deposited LAO, STO, YSZ and LSMO films by PLD at room temperature on (001)-oriented TiO$_2$-terminated STO substrates (Figure S1, Supporting Information), which resulted in atomically smooth film surfaces (see inset Figure 1a). For comparison, all the films were also deposited on (001)-oriented LAO and (LaAlO$_3$)$_{0.3}$(SrAl$_{0.5}$Ta$_{0.5}$O$_3$)$_{0.7}$ (LSAT) substrates simultaneously. The used growth conditions are similar to the ones commonly used in the literature except for the ambient deposition temperature (Table S1, Supporting Information). At this temperature the deposited films are amorphous and any oxygen exchange between the substrate and the background gas is significantly suppressed.

As shown in Figure 1a, the LAO/STO, STO/STO, and YSZ/STO heterostructures with amorphous top layers, exhibit metallic conducting behavior. In contrast, each of the heterostructures grown on both LAO and LSAT substrates is highly insulating. These results indicate that the deposited films are insulating and that the metallic conduction in STO-based heterostructures is closely related to the STO substrate. However, it should be mentioned that no conductivity is observed in any amorphous LSMO/STO heterostructures, as discussed below. For the conductive samples based on STO substrates, Hall-effect measurements show that the charge carriers at the interface are electron type, *i.e.* n-type. The obtained sheet carrier density, $n_s$, is nearly constant in the temperature range of 100-300 K with a value of (0.8-1.2)×10$^{14}$ cm$^{-2}$ (Figure 1b). As the temperature decreases below 100 K, the sheet carrier decreases, and the Hall resistance becomes non-linear with respect to magnetic fields at temperatures below 30 K (Figure S2, Supporting Information). At 2 K, $n_s$ reaches a value around $5\times10^{13}$ cm$^{-2}$. The electron mobility, $\mu_s$, increases upon cooling, and is around 200 cm$^2$V$^{-1}$s$^{-1}$ at 2 K (Figure 1c). Interestingly, the carrier density for the conductive amorphous STO-based heterostructures is of the same order of magnitude as those reported for the STO-based heterostructures with crystalline overlayers deposited at high temperatures.[4-6,12,13,15]

The interfacial conductivity of the amorphous STO-based heterostructures exhibits strong dependence on the $PO_2$ during film growth. As shown in Figures 2a and b, for samples with film thickness around 30 nm, the heterointerfaces grown at $PO_2 > 1\times10^{-2}$ mbar are all insulating, with $R_s$ equal to that of the bare STO substrates ($R_s > 10^9$ ohms per square, measurement limit). Upon decreasing the pressure below $1\times10^{-2}$ mbar, the heterointerfaces of LAO/STO, STO/STO, and YSZ/STO turn conductive (See also Figure S3 in Supporting Information), whereas LSMO/STO remains insulating. It is further found that the temperature dependence of the conducting behavior in the amorphous LAO/STO samples (Figure S3a, Supporting Information) exhibits similar characteristics with those reported in crystalline LAO/STO samples grown at $PO_2$ below $1\times10^{-2}$ mbar.[6] The prime differences are found in the onset temperature of a semiconducting behavior, which seems to be lowered with decreasing $PO_2$. On the other hand, a simply metallic behavior is observed in the samples of both STO/STO and YSZ/STO (Figure S3b, Supporting Information) grown at $PO_2 \leq 1\times10^{-3}$ mbar.

More remarkably, an interfacial metal-insulator transition that depends on the thickness of the amorphous film is observed in each kind of the heterostructures of LAO/STO, STO/STO, and YSZ/STO, when films are deposited at $PO_2 \approx 1\times10^{-6}$ mbar. As shown in Figures 2c and d, the heterointerface is insulating at $t < 1.8$ nm for LAO/STO, $t < 2.2$ nm for YSZ/STO, and $t < 2.8$ nm for STO/STO, respectively. Astonishingly, these heterointerfaces abruptly become metallic upon further increase of the film thickness. Note that the electronic properties remain constant with further increase of the film thickness. The critical thickness for the occurrence of conductivity, which depends on the deposited material, may also increase upon increasing the $PO_2$ as observed for our LAO/STO samples (Figure S4, Supporting Information).

To determine the origin of the interfacial conductivity in our STO-based heterostructures, *in-situ* X-ray photoelectron spectroscopy (XPS) measurements were performed. Figures 3a and b show the Ti $2p_{3/2}$ spectra for different film thicknesses of LAO/STO and YSZ/STO samples, respectively, grown at $PO_2 \approx 1\times10^{-6}$ mbar. For STO/STO samples, the broad Ti $2p_{3/2}$ core-level spectra of the amorphous STO film obscure the details of the STO substrate. As shown in Figures 3a and b, no clear $Ti^{3+}$ signal in the $2p_{3/2}$ core-level spectra could be detected in the bare STO substrate, as expected. However, finite amount of $Ti^{3+}$ is already present in the insulating samples, even at $t = 0.4$ nm. This suggests the formation of defects in STO, either by intermixing or by oxygen vacancies.[18] The amount of $Ti^{3+}$ increases with increasing film thickness, as shown in Figure 3c. This result, which is difficult to explain by intermixing, may indicate the evolution of oxygen vacancies in STO substrates upon film deposition. Note that the amount of $Ti^{3+}$ is stable over time after film deposition (inset Figure S5b, Supporting Information), which strongly indicates that the formation of oxygen vacancies only occurs during film deposition. Additionally, in contrast to the negligible difference in the measured conductivity, a distinct higher concentration of $Ti^{3+}$ is present in YSZ/STO compared to LAO/STO (Figure 3c). This indicates that the concentration of $Ti^{3+}$ is not directly proportional to the conductivity as the critical thickness of LAO/STO is lower than that of YSZ/STO. For $Ti^{3+}$ depth profiling, angle-resolved XPS measurements were performed. No clear angle dependence of the $Ti^{3+}$ signal is observed, independent of the film thickness (Figures S5a and b, Supporting Information). This indicates that $Ti^{3+}$ extend several nanometers deep into the STO substrate, though it is expected to be confined in the vicinity of the interface. These results are different from the results of crystalline LAO/STO, where a clear angle dependence of the $Ti^{3+}$ signal has been observed.[19] Additionally, the effect of $PO_2$ on the concentration of $Ti^{3+}$ was investigated in YSZ/STO. Only small differences in the $Ti^{3+}$ signal are observed between heterostructures grown at $1\times10^{-6}$ mbar $\leq PO_2 \leq 1\times10^{-3}$ mbar (Figure S5c, Supporting Information). This is consistent with the transport measurements as shown in Figures 2a and b. In short, the XPS results suggest that the interfacial conductivity in our STO-based heterostructures should be mainly ascribed to the oxygen vacancies on the STO substrate side. This conclusion is also consistent with the clearly reduced amount of $Ti^{3+}$ in the LSMO/STO heterostructure with a 2 nm film grown at $PO_2 \approx 1\times10^{-6}$ mbar (Figure 3c), where the effect of heavy charging during XPS measurements constrains the XPS signals for a higher film thickness. In addition, the conductivity in these amorphous STO-based heterostructures can be removed by annealing in 0.5-1 bar pure $O_2$ at 150-300 ºC. For example, annealing a 2.1 nm LAO/STO sample even at 150 ºC for 1.5 hours removed the conductivity completely and reduced the $Ti^{3+}$ signal almost to zero (Figure 3c and Figure S6, Supporting Information). This further supports that oxygen vacancies account for the interfacial conductivity in our amorphous STO-based heterostructures.

Based on aforementioned results, besides the deposited materials, the oxygen pressure during film growth turns out to be the main factor that controls the interfacial conduction. In a PLD process, the pressure determines the expansion dynamics of the plume and the composition of the plasma.[20] In the pressure range of $PO_2<1\times10^{-2}$ mbar, the PLD plasma expands freely, which consist of a large fraction of energetic atomic neutrals and a small fraction (in the range of 1%-5%[21]) of ions, with energies between tens and hundreds of eV.[20,21] Note that the interfacial conduction is mainly observed in this oxygen pressure range. Increasing the pressure to $PO_2>1\times10^{-2}$ mbar generally results in collisions between the plasma plume and the background gas, leading to a much lower energy for the plasma species.[20] More importantly, in an oxygen background at this high pressure, the plasma species will be oxidized before arriving at the substrate.[20,21] Note that the interfacial conductivity becomes negligible in this regime. Therefore, the $PO_2$ dependence of the plasma properties and the interfacial conduction exhibit close correspondence. Consequently, two scenarios should be considered for the interfacial conductivity: first, the impinging of the high energetic plasma species into the substrate, and second, the chemical reactivity of the plasma species with respect to the STO substrate.

The first scenario, defect creation at the STO substrate surface due to the high energy of the impinging plasma species[15-17] is similar to the $Ar^+$ irradiation induced conductivity in STO.[22] When ions impinge on the substrate surface, the energy loss is typically of several hundreds of eV per nm. The penetration length of the PLD plasma should therefore be much lower than 1 nm, comparable to $Ar^+$ irradiation at 100 eV.[22] When the thickness of the grown film is larger than 1 nm, the bombarding effect on STO substrates should become negligible. Therefore, the sputtering effect, with possible cation intermixing involved, is especially a concern during deposition of the first monolayer of the film. As interfacial conductivity is observed for larger film thicknesses, it is expected that the sputtering effect has a much smaller contribution to the conductivity than the chemical effect described below. Moreover, the sputtering scenario is not compatible with the increase of the density of oxygen vacancies in STO substrates upon increasing film thickness, as indicated in Figure 3c. Additionally, with the assumption of a similar kinetic distribution of plasma species when ablating different oxides, a sputtering scenario is not consistent with the insulating interfaces between STO substrates and amorphous LSMO films, which are nearly independent of film thickness and $P$O$_2$.

In the second scenario, taking the chemical reactivity of the plasma into account, the deposited species react with the oxygen ions (O$^{2-}$) present in the STO substrate lattice. Interestingly, the defect formation energy of the TiO$_2$-terminated STO surface (5.94 eV[23]), which is considerably smaller than that of the bulk,[23] is comparable to the bond dissociation energy of oxygen molecules (5.11 eV[21]). In this case, besides the oxygen source from the target and the background oxygen gas, the oxygen ions in STO substrates most likely diffuse outward to oxidize the reactive plasma species absorbed on the STO substrate surface, as schematically depicted in Figure 4. Similar to this, the outward diffusion of oxygen ions from the STO substrate has been previously observed during the growth of reactive metals of Ti and Y films by molecular beam epitaxy under ultrahigh vacuum.[24] Therefore, the oxygen vacancies and the resulting interfacial conductivity may result from the redox reactions at the interface by reducing the STO substrate surface to oxidize the oxygen-deficient overlayer.

The chemical interactions at the interface between a metal and the TiO$_2$ or STO substrate surface are controlled not only by the thermodynamic stability of the metal oxide, but also by the space charges at the metal/oxide interface, which is determined by the interface electronic configuration, i.e. the relative Fermi level of the metal and that of the TiO$_2$ or STO before contact.[25] An interfacial redox reaction can occur at room temperature when the heat of metal oxide formation per mole of oxygen, $\Delta H_f^O$, is lower than -250 kJ/(mol O) and the work function of the metals, φ, is in the range of 3.75 eV<φ <5.0 eV.[25]

Interestingly, for our film growth with conducting interfaces, one of the main neutral species in each freely-expanding plume (Al for the LAO plume, Ti for the STO plume, and Zr for the YSZ plume) follows the above criterion. On the other hand, Mn, the main atomic composition of the corresponding LSMO plasma, locates on the border region for the occurrence/nonoccurrence of redox reactions on the $TiO_2$ surface at room temperature.[25] Though the redox reactions at the interface between complex oxides are much more complicated than those at the metal/oxide interfaces, the lack of redox reactions at the interface of our amorphous LSMO/STO samples should explain their insulating interfaces. It should be mentioned that a possible sputtering effect, which is especially a concern during the deposition of the first monolayer of the films, cannot enable conductivity directly at larger film thicknesses, but could facilitate or enhance the oxygen ions outward diffusion leading to interfacial conductivity.

In summary, we have shown clear evidence that the chemical composition of the film growth plasma affects the conduction of amorphous STO-based heterostructures by introducing oxygen vacancies near the STO substrate surface. Our results indicate that, besides electronic interactions, chemical reactions at the interface play an important role in determining interfacial conductivity in STO-based heterostructures. This provides new opportunities to design metallic or insulating interfaces in STO-based heterostructures. The easy room temperature fabrication of highly conductive interfaces of complex oxide heterostructures is promising for the application in both oxide electronics and thermoelectric oxides.

**METHODS**

**Sample growth**. The LAO, STO, YSZ and LSMO films were grown under different oxygen pressure of $1\times10^{-6}$ to 1 mbar by PLD using a KrF laser ($\lambda$=248 nm) with a repetition rate of 1 Hz and laser fluence of 1.0-2.0 Jcm$^{-2}$ in both Risø and Twente laboratories. The target-substrate distance was fixed at 4.5 cm. Commercial LAO, STO, and YSZ (with 9 mol. % nominal yttria content) single crystals and sintered LSMO ceramics were used as targets. Substrates of (001)-oriented single crystalline STO, LAO, and LSAT with a size of 5×5×0.5 mm$^3$ were used. For the STO substrates, a singly $TiO_2$-termination was achieved by chemical etching.[1,2] Films deposited at room temperature showed amorphous structure, which was confirmed by in-situ high pressure RHEED. All films showed flat surfaces with a root-mean-square roughness less than 0.5 nm measured by atomic force microscopy (AFM). The film thickness was determined by AFM through patterning the samples prior to deposition, which was also checked by X-ray reflectivity measurements.

**Electronic transport measurement**. The sheet resistance and carrier density of the buried interface were measured using a 4-probe Van der Pauw method with ultrasonically wire-bonded aluminum wires as electrodes. The temperature dependent electrical transport and Hall-effect measurements were performed in a Quantum Design physical properties measurement system (PPMS) in the temperature range from 300 K down to 2 K with magnetic fields up to 14 T. To determine the exact carrier density, Hall-bar patterned samples were also measured, which were prepared through a mechanical mask.

***In-situ* X-ray photoelectron spectroscopy (XPS) measurement**. For the study using angular resolved XPS (by the Twente group), a series of samples were first grown by PLD. Subsequently, the samples were transferred to the XPS chamber while keeping them under high vacuum (below $1\times10^{-9}$ mbar). The XPS chamber (Omicron Nanotechnology GmbH) had a base pressure below $1\times10^{-10}$ mbar. The measurements were done using a monochromatic Al Kα (XM 1000) X-ray source and an EA 125 electron energy analyzer. All spectra were acquired in the Constant Analyzer Energy (CAE) mode. The escape angle of the electrons was varied by rotation of the sample between 15 to 80 degrees with respect to the analyzer normal. A CN 10 charge neutralizer system was used to overcome the charging effect in the LSMO/STO heterostructures. For each measurement the filament current, emission current and beam energy were optimized to minimize the full width at half maximum (FWHM) of the peaks. For analyzing the Ti $2p_{3/2}$ peaks, a Shirley background was subtracted and the spectra were normalized on the total area below the Ti peaks ([Ti] = [Ti4+] + [Ti3+] = 100%).

**Acknowledgement.** We thank C. R. H. Bahl, N. H. Andersen, F. B. Saxild, J. Geyti, K. V. Hansen, N. Bonanos, T. Y. Zhao, W. W. Gao, A.Z. Jin, D. V. Christensen, F. Trier, C. Hougaard, and A. Jørgensen for their valuable help. G.R. thanks the financial support by the Netherlands Organization for Scientific Research (NWO) through a VIDI grant. J.R.S. and B.G.S. thank the support of the National

**Supporting Information Available**. This material is available free of charge via the Internet at http://pubs.acs.org.

**FIGURE CAPTIONS**.

**Figure 1.** (a) The temperature dependent sheet resistance, $R_s$, of the LAO/STO, STO/STO, and YSZ/STO heterostructures with about 8 nm amorphous capping films grown at $PO_2 \approx 1\times10^{-6}$ mbar. The inset shows a 5μm×5μm atomic force microscopy image of the STO/STO sample with regular terraces of ~0.4 nm in height. (b) and (c) The sheet carrier density, $n_s$, and electron mobility, $\mu_s$, versus temperature, respectively.

**Figure 2.** (a) and (b) The sheet resistance, $R_s$, and carrier density, $n_s$, versus oxygen pressure, $PO_2$, respectively, for the LAO/STO, STO/STO, YSZ/STO and LSMO/STO samples with a film thickness, $t$, of around 30 nm. (c) and (d) $R_s$ and $n_s$ versus $t$, respectively, for all the STO-based heterostructures grown at $PO_2 \approx 1\times10^{-6}$ mbar. The results were measured at $T$=300 K.

**Figure 3.** (a) and (b) The Ti $2p_{3/2}$ XPS spectra for different film thicknesses of amorphous LAO/STO and YSZ/STO, respectively. The insets show a close-up of the Ti$^{3+}$ peak. All films were grown at $PO_2 \approx 1\times10^{-6}$ mbar. The spectra were measured at an emission angle of 80°. (c) The film thickness dependent percentage of Ti$^{3+}$ for LAO/STO, YSZ/STO and LSMO/STO samples, where [Ti$^{3+}$] + [Ti$^{4+}$] = 100%. Annealing the 2.1 nm LAO/STO sample at 150 °C for 1.5 hours in 0.6 bar pure O$_2$ reduced the Ti$^{3+}$ signal almost to zero.

**Figure 4.** Schematic representation for the oxygen ions outward diffusion induced interfacial conduction in STO-based heterostructures during growth of the oxide films at room temperature.

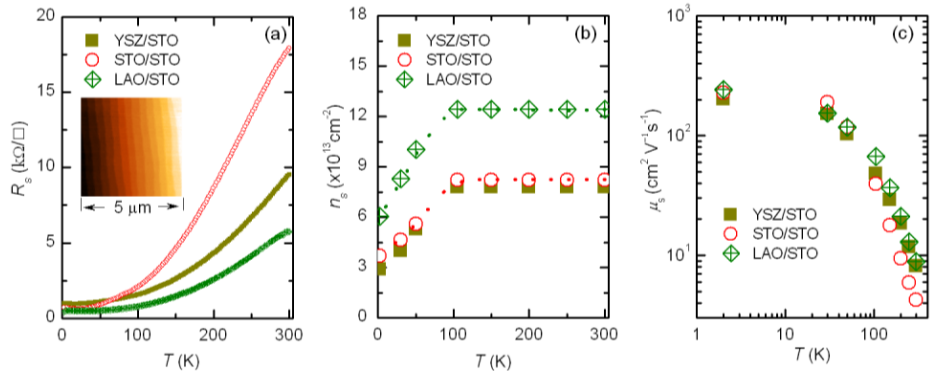

Fig. 1. *Y. Z. Chen et al.*

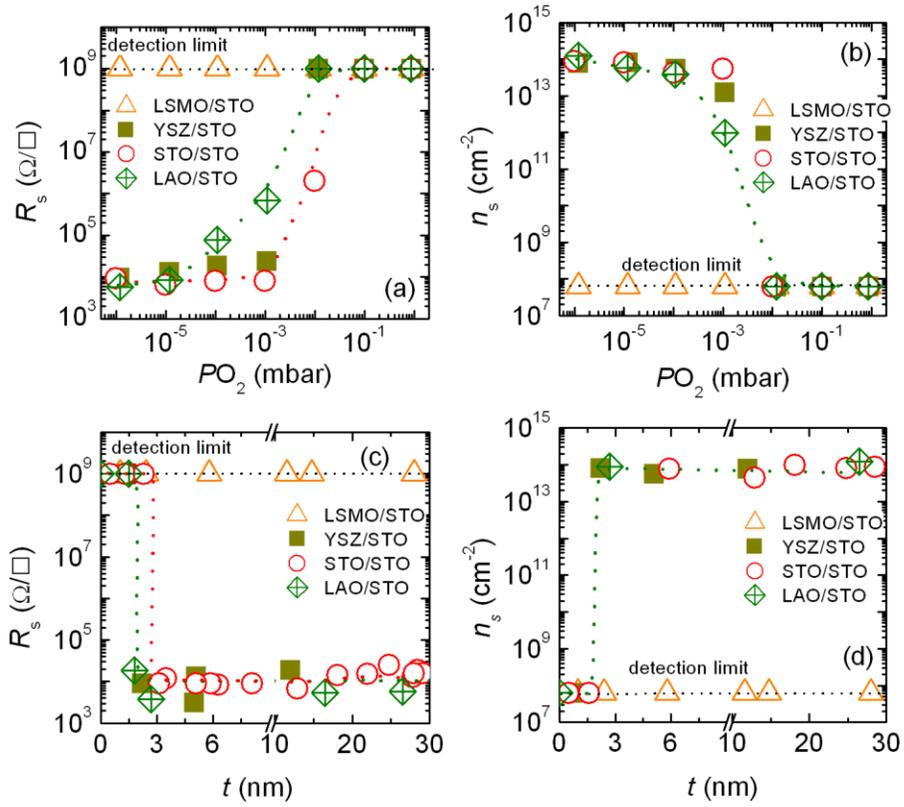

Fig.2 Y. Z. Chen *et al*.

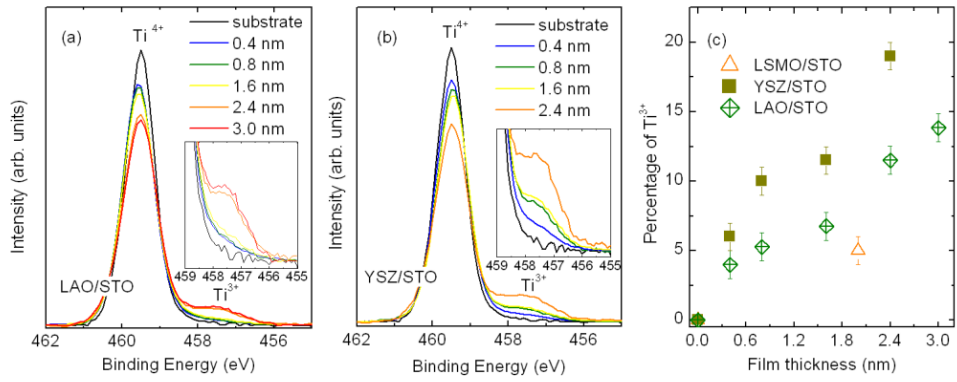

Fig. 3. Y. Z. Chen *et al*.

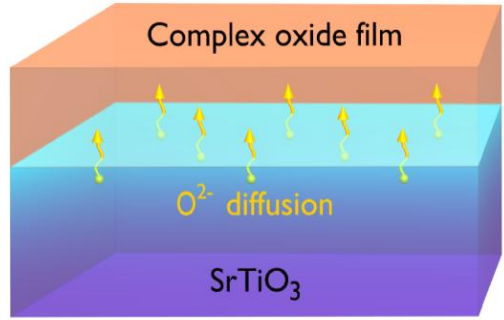

Fig.4 Y. Z. Chen *et al.*